\documentclass[final,3p,times]{elsarticle}

\usepackage{amssymb}
\usepackage{amsmath}
\usepackage{float}

\journal{Journal of Subatomic Particles and Cosmology}

\begin{document}

\begin{frontmatter}

\title{Measurements of local polarization of $\Lambda$ hyperons \\
       with event shape engineering in Au+Au collisions at RHIC-STAR}

\author[aaa]{Taiki Kondo}
\makeatletter
\g@addto@macro\elsauthors{~for the STAR Collaboration}
\g@addto@macro\elsprelimauthors{~for the STAR Collaboration}
\makeatother

\affiliation[aaa]{organization={University of Tsukuba},
             addressline={1-1-1 Tennodai},
             city={Tsukuba},
             postcode={305-8577},
             state={Ibaraki},
             country={Japan}}

\begin{abstract}
Anisotropic flow is expected to generate local vorticities along the beam direction, which in turn lead to local polarization of produced particles.
Although the local polarization of hyperons has already been observed at RHIC and the LHC, its mechanism is not yet understood: model calculations indicate that the shear-induced and vorticity-induced contributions can have opposite signs and may substantially cancel, so that the resulting polarization depends sensitively on the theoretical implementation.
Since both are driven by the anisotropic flow, we employed the Event Shape Engineering (ESE) method, which selects events with different flow magnitudes within a narrow centrality range, to measure the local polarization as a function of the elliptic flow at a fixed multiplicity.
We report the measurements of the elliptic flow ($v_2$) and the local polarization ($P_{z,2}$) of $\Lambda$ and $\bar{\Lambda}$ as a function of the reduced flow vector magnitude $q_2$ used for the ESE event selection, in Au+Au collisions at $\sqrt{s_{\rm NN}} = 19.6~{\rm GeV}$ from the STAR experiment.
The $v_2$ shows a clear positive dependence on $q_2$, while $P_{z,2}$ indicates a weak positive dependence on $q_2$ and $v_2$.
This is the first measurement to directly investigate the relation between anisotropic flow and local polarization, providing new insight into the origin of local polarization.

\end{abstract}


\end{frontmatter}


\section{Introduction}
In heavy-ion collisions, anisotropic flow, such as elliptic flow, is expected to generate local vorticities along the beam direction, which lead to the spin polarization of produced particles, called local polarization.
This local polarization of hyperons has already been observed at RHIC \cite{AuAu200,Isobar} and the LHC \cite{LHC}, but its mechanism is still under debate.
Model calculations indicate that the shear-induced and vorticity-induced contributions can have opposite signs and may substantially cancel, with the resulting polarization depending sensitively on the theoretical implementation \cite{Wu,Fu,Alzhrani,Niida}.
Therefore, we applied the Event Shape Engineering (ESE) method \cite{Schukraft}, which selects events with different magnitudes of the final-state flow within a narrow centrality bin, to our polarization measurements.
In this study, we measured the elliptic flow and local polarization of $\Lambda$ and $\bar{\Lambda}$ hyperons in Au+Au collisions at $\sqrt{s_{\rm NN}} = 19.6~{\rm GeV}$ from the STAR Beam Energy Scan Phase-II program.
Using the ESE method, we investigate the relation between the elliptic flow and the local polarization.

\section{Analysis details}

\begin{figure}[H]
  \centering
  \includegraphics[width=0.6\textwidth]{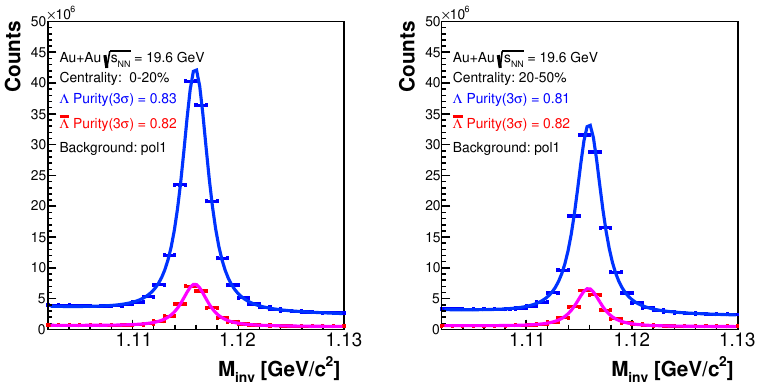}
  \caption{
      The invariant mass distributions of the proton-pion candidates ($p\pi^{-}$ for $\Lambda$ and $\bar{p}\pi^{+}$ for $\bar{\Lambda}$) in the 0--20\% (left) and 20--50\% (right) centrality bins in Au+Au collisions at $\sqrt{s_{\rm NN}} = 19.6~{\rm GeV}$.
      The distributions were fitted with a Student's t-function.
      }
  \label{fig:Minv}
\end{figure}

In these measurements, the Time Projection Chamber (TPC) was used for particle identification (PID) and track reconstruction.
The Time of Flight (ToF) detector was also used to improve PID accuracy.
The Event Plane Detector (EPD) was utilized for event-plane determination at forward and backward rapidities.
For the $\Lambda$ and $\bar{\Lambda}$ reconstruction, we first identified daughter protons and pions using TPC and ToF information.
The $\Lambda$ and $\bar{\Lambda}$ were then reconstructed based on their decay topology.
The invariant mass distributions of the proton-pion candidates in the 0--20\% and 20--50\% centrality bins are shown in Figure \ref{fig:Minv}.
Daughter pairs within a $3\sigma$ mass window around the invariant mass peak were selected for further analysis.

\begin{figure}[H]
  \centering
  \begin{minipage}[t]{0.49\textwidth}
    \centering
    \includegraphics[width=\textwidth]{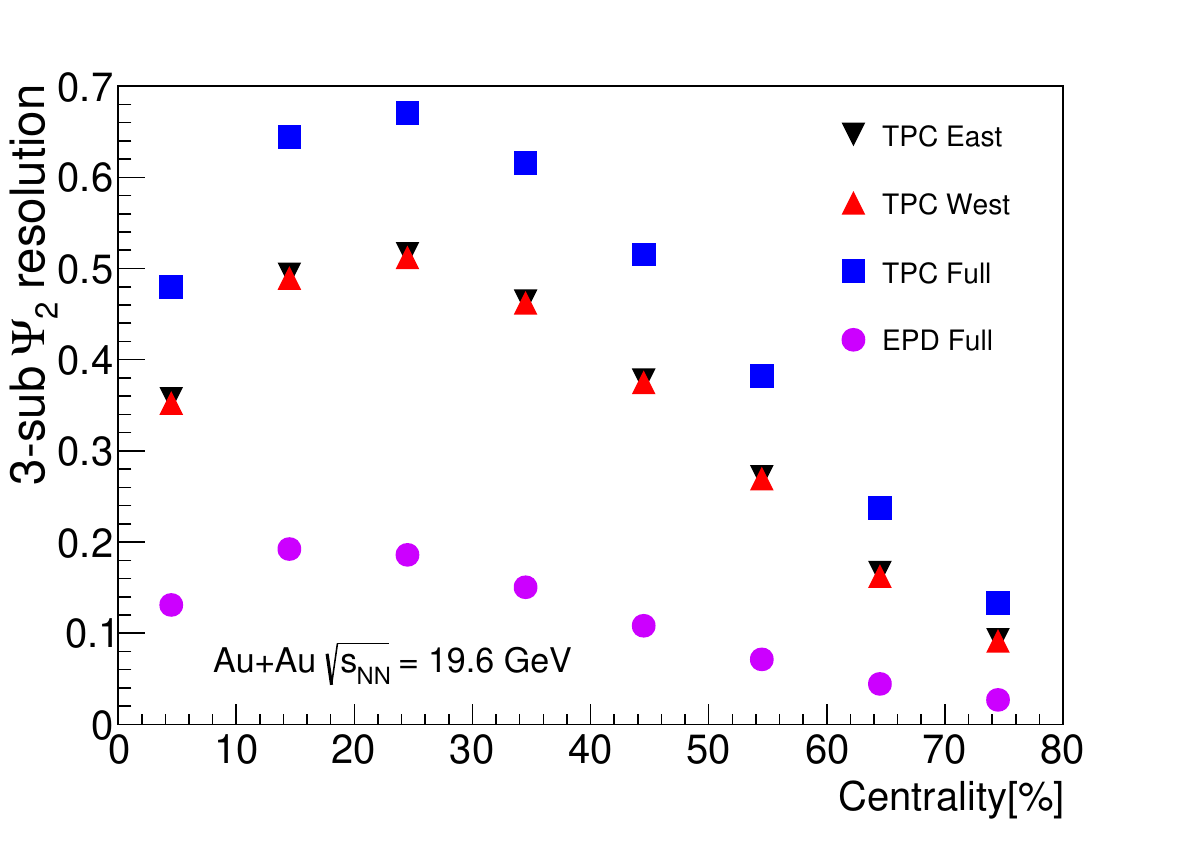}
    \caption{
      The event-plane resolution as a function of centrality for the TPC-Full, TPC-Sub, and EPD-Full sub-events, which were estimated using the 3-sub event method in Au+Au collisions at $\sqrt{s_{\rm NN}} = 19.6~{\rm GeV}$.}
    \label{fig:Res2}
  \end{minipage}
  \hfill
  \begin{minipage}[t]{0.45\textwidth}
    \centering
    \includegraphics[width=\textwidth]{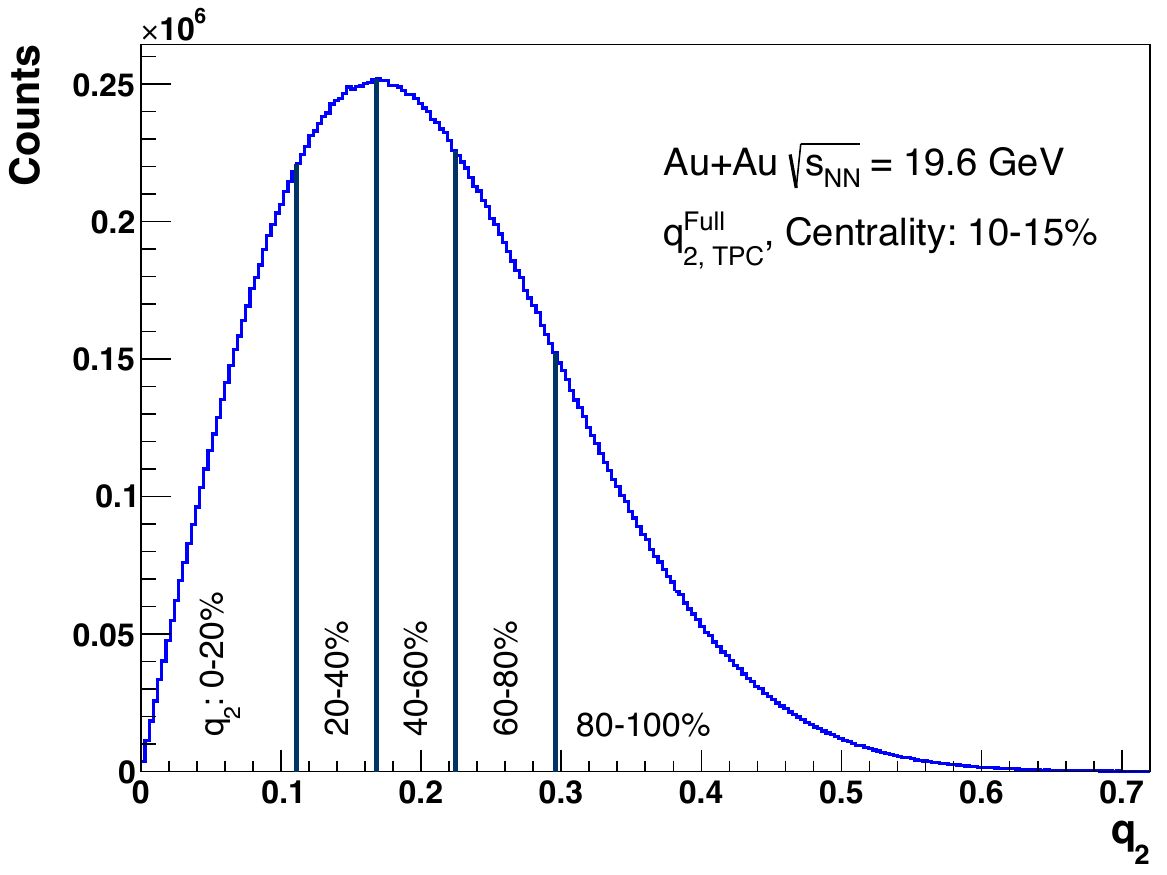}
    \caption{
      The $q_2$ distribution for the TPC-Full sub-event in the 10--15\% centrality bin in Au+Au collisions at $\sqrt{s_{\rm NN}} = 19.6~{\rm GeV}$.}
      \label{fig:q2}
  \end{minipage}
\end{figure}

We used three types of sub-events, TPC-Full ($|\eta| < 1.5$), TPC-Sub, consisting of TPC-East ($-1.5 < \eta < -0.1$) and TPC-West ($0.1 < \eta < 1.5$), and EPD-Full ($2.14 < |\eta| < 5.09$), for the event-plane determination and event selection.
The event-plane resolution is shown as a function of centrality for each sub-event in Figure \ref{fig:Res2}.
For the ESE procedure, the reduced flow vector magnitude $q_2 = |Q_2|/\sqrt{\sum p_{\rm T}}$ is calculated from charged tracks with $0.15 < p_{\rm T} < 2.0$~GeV/$c$, where the upper limit suppresses non-flow contributions from jet fragmentation.
Here $Q_2$ is the second-order flow vector defined as $Q_2 = \left( \sum_i p_{{\rm T},i}\cos(2\phi_i),\ \sum_i p_{{\rm T},i}\sin(2\phi_i) \right)$,
where the sum runs over all charged particles $i$ used in the flow vector determination, and $\phi_i$ is the azimuthal angle of particle $i$.
The normalization by $\sqrt{\sum p_{\rm T}}$ removes the trivial growth of $|Q_2|$ with the event multiplicity, so that $q_2$ can be compared among events within a centrality bin.
A larger $q_2$ value corresponds to events with a larger elliptic flow.
The transverse-momentum weighting enhances the correlation between $q_2$ and the elliptic flow $v_2$, thereby improving the sensitivity of the event-shape selection.
The $q_2$ distribution for the TPC-Full sub-event in the 10--15\% centrality bin is shown in Figure \ref{fig:q2}.
Events were divided into five groups with equal statistics based on the $q_2$ distributions.
To minimize correlations of the $q_2$ selection with the event multiplicity, the selection was performed within 5\% centrality bins, with additional fine multiplicity binning for the 0--10\% centrality range.

The elliptic flow $v_2$ is given by:
\begin{equation}
  v_2 = \frac{\langle\cos [2(\phi_{\Lambda}-\Psi_{2})]\rangle}{\mathrm{Res}\{\Psi_2\}},
  \label{eq:v2}
\end{equation}
where $\phi_{\Lambda}$ is the azimuthal angle of the $\Lambda$, $\Psi_2$ is the second-order event-plane angle, and $\mathrm{Res}\{\Psi_2\}$ is the event-plane resolution.
The event-plane method is used for the polarization measurement \cite{Pz}.
The polarization component $P_{z,2}$ is evaluated as:
\begin{equation}
  P_{z,2} = \frac{1}{\alpha_\Lambda \langle \cos^2 \theta_p^*\rangle}
  \frac{\langle\cos\theta^*_p \sin(2(\phi_\Lambda-\Psi_2))\rangle}{\mathrm{Res}\{\Psi_2\}},
  \label{eq:Pz}
\end{equation}
where $\alpha_\Lambda$ is the $\Lambda$ decay parameter $(\alpha_{\Lambda} = -\alpha_{\bar{\Lambda}} = 0.747 \pm 0.010)$, and $\theta_p^*$ is the polar angle of the daughter proton in the $\Lambda$ rest frame.
Both $v_2$ and $P_{z,2}$ were measured separately for each $q_2$ class.
To thoroughly check the systematics of the measurement and possible non-flow effects, we tested four combinations of event plane and $q_2$:
1.~($\Psi_{2~\rm TPC}^{\rm Sub},\,q_{2~\rm TPC}^{\rm Sub}$),
2.~($\Psi_{2~\rm TPC}^{\rm Full},\,q_{2~\rm TPC}^{\rm Full}$),
3.~($\Psi_{2~\rm EPD}^{\rm Full},\,q_{2~\rm TPC}^{\rm Sub}$),
4.~($\Psi_{2~\rm EPD}^{\rm Full},\,q_{2~\rm TPC}^{\rm Full}$).
For $\Psi_{2~\rm TPC}^{\rm Sub}$ and $q_{2~\rm TPC}^{\rm Sub}$, the sub-event in the opposite $\eta$ hemisphere to the $\Lambda$ is used.
The $\Lambda$ daughter tracks are explicitly excluded from the $Q$-vector calculations to avoid self-correlation effects.
The measured $v_2$ and $P_{z,2}$ within the $3\sigma$ mass window include contributions from the combinatorial background under the $\Lambda$ peak.
The purity $P$ is obtained by fitting the invariant-mass distribution with a Student's t-function for the signal and a first-order polynomial for the background in each centrality bin.
The signal values are then extracted as $S = [S_{\rm obs} - (1-P)S_{\rm bkg}]/P$, where $S_{\rm obs}$ is the value measured in the signal window and $S_{\rm bkg}$ is estimated from the sideband regions, $1.090 < M_{\rm inv} < 1.105$ and $1.125 < M_{\rm inv} < 1.140$ GeV/$c^2$.
The systematic uncertainties were evaluated by varying the topological selection criteria for the $\Lambda$ reconstruction, the invariant-mass window, the background estimation, and the $\Lambda$ decay parameter.
They are largely correlated among the $q_2$ classes and therefore affect the $q_2$ and $v_2$ dependences less than the absolute values.
The significances quoted below are based on statistical uncertainties only.

\section{Results and discussion}

\begin{figure}[H]
  \centering
  \includegraphics[width=0.80\textwidth]{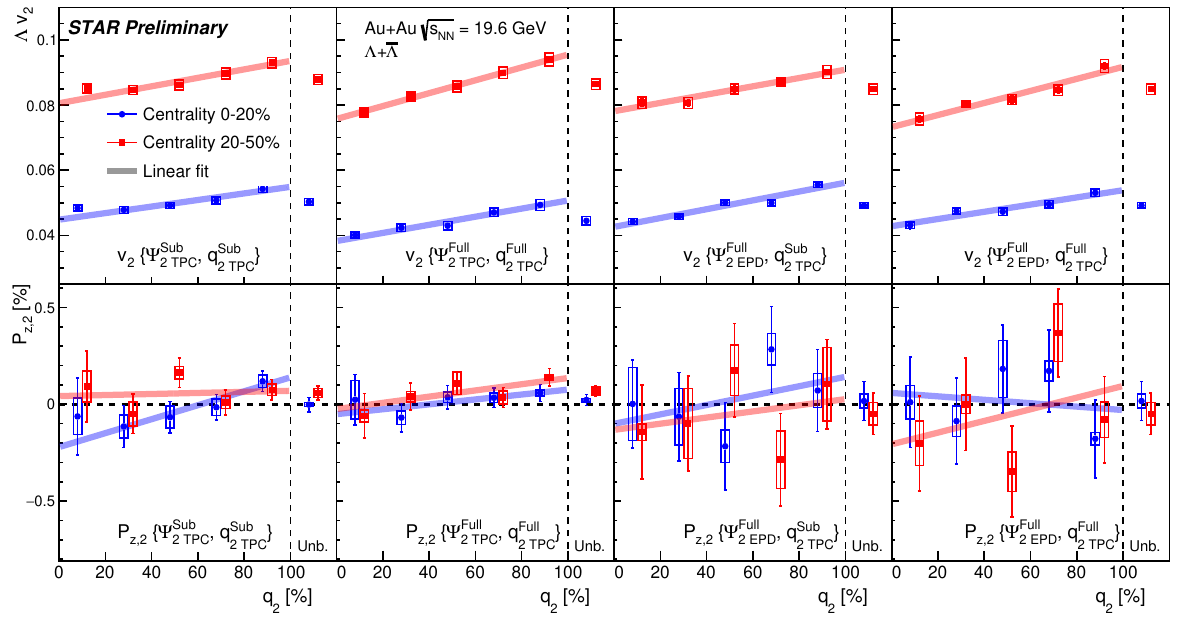}
  \caption{
      The $q_2$ dependence of $v_2$ (top panels) and $P_{z,2}$ (bottom panels) averaged for $\Lambda$ and $\bar{\Lambda}$ for four combinations of event plane $\Psi_2$ and $q_2$ in Au+Au collisions at $\sqrt{s_{\rm NN}} = 19.6~{\rm GeV}$.
      The most-right data points in each panel denoted by ``Unb.'' represent unbiased signals.}
  \label{fig:v2Pz}
\end{figure}

The $q_2$ dependence of $v_2$ and $P_{z,2}$ for $\Lambda + \bar{\Lambda}$ is shown in Figure \ref{fig:v2Pz} for four combinations of event plane and $q_2$.
The $\Lambda$ $v_2$ shows a clear positive dependence on $q_2$ across all combinations.
The $v_2$ results are consistent in all approaches.
Note that the combination ($\Psi_{2~\rm EPD}^{\rm Full}, q_{2~\rm TPC}^{\rm Sub}$) determines $\Psi_2$ and $q_2$ from different detectors covering different $\eta$ regions, so that they share no particles.
The consistency among the different combinations therefore indicates that possible biases associated with shared particles are small within the present uncertainties.
Given its higher event-plane resolution and $q_2$ sensitivity, the combination utilizing the TPC-Full event plane and TPC-Full $q_2$ provides the best statistical sensitivity, while the other configurations serve as important checks of possible shared-particle and non-flow effects.
For the polarization measurements, $P_{z,2}$ measured with the TPC event plane and TPC $q_2$ exhibits a slight positive dependence on $q_2$, while $P_{z,2}$ measured with $\Psi_{2~\rm EPD}^{\rm Full}$ is less significant due to the lower resolution of the event plane.

\begin{figure}[H]
  \centering
  \includegraphics[width=0.55\textwidth]{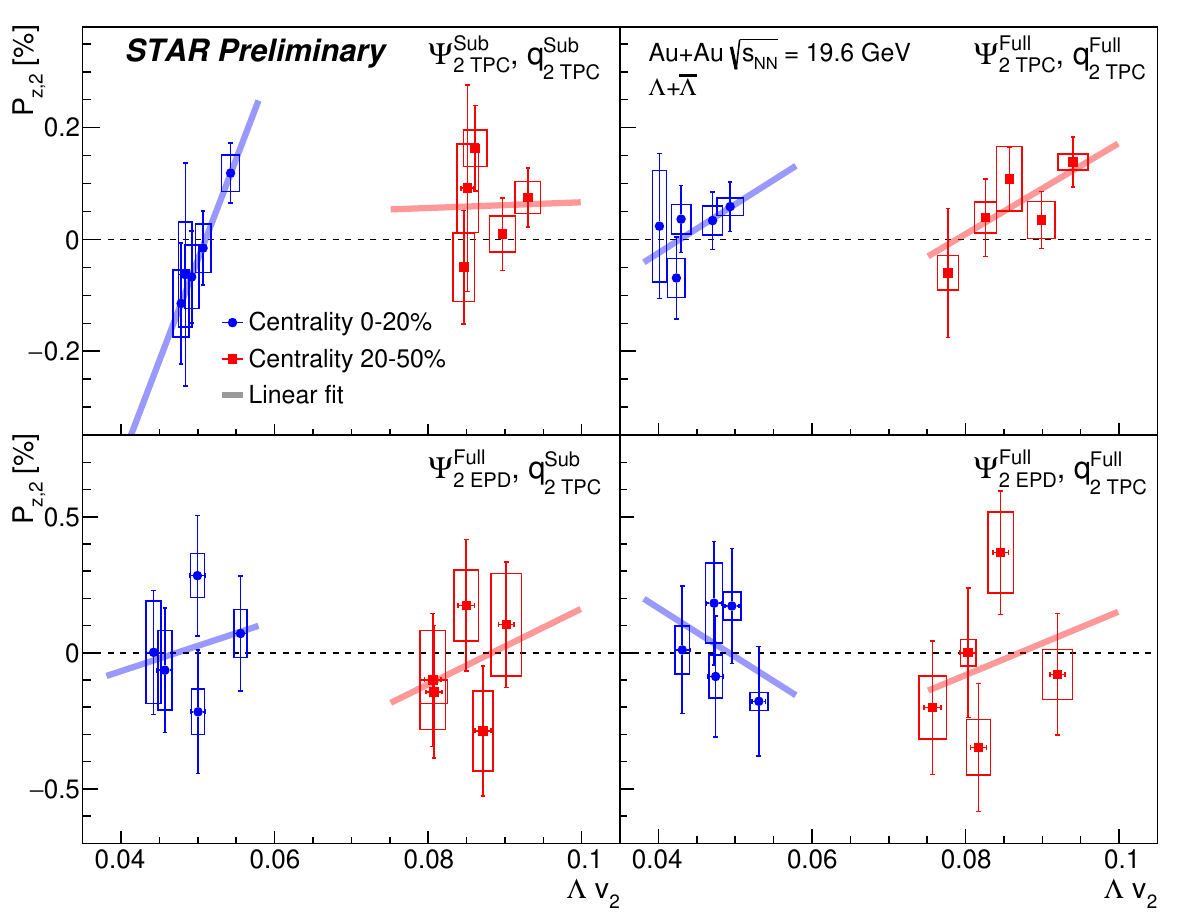}
  \caption{
      The $P_{z,2}$ signals averaged for $\Lambda$ and $\bar{\Lambda}$ as a function of $v_2$ for four combinations in Au+Au collisions at $\sqrt{s_{\rm NN}} = 19.6~{\rm GeV}$.}
  \label{fig:v2vsPz}
\end{figure}

The $v_2$ dependence of $P_{z,2}$ for $\Lambda + \bar{\Lambda}$ is shown in Figure \ref{fig:v2vsPz}.
For the $P_{z,2}$ signals extracted using the TPC event plane and TPC $q_2$, a positive slope is observed with a significance of approximately $2.5\sigma$, suggesting a possible correlation between the elliptic flow and the local polarization.
The fitted intercept is negative at approximately the $2.5\sigma$ level for this combination.
While a non-zero intercept may indicate contributions not directly correlated with the elliptic flow, this interpretation should be treated with caution, because the intercept is obtained by extrapolating to $v_2 = 0$, far outside the $v_2$ range covered by the ESE selection, where the fit is not constrained.
Indeed, one of the four combinations yields a positive intercept, as seen in Figure \ref{fig:v2vsPz}.

\section{Summary and outlook}

Elliptic flow and local polarization of $\Lambda$ and $\bar{\Lambda}$ have been studied in Au+Au collisions at $\sqrt{s_{\rm NN}} = 19.6~{\rm GeV}$ with the ESE method using the STAR detector at RHIC.
This is the first measurement to investigate the relation between elliptic flow and local polarization.
The polarization signals measured with the TPC event plane and TPC $q_2$ indicate a slight positive slope as a function of both the charged-particle $q_2$ and the $\Lambda$ $v_2$.
The observed data suggest that anisotropic flow may contribute to local polarization.
The intercept of the linear fit to $P_{z,2}$ as a function of $v_2$ may carry information on contributions not correlated with the elliptic flow, although it is obtained by extrapolation to $v_2 = 0$ and is therefore not yet conclusive.
As an outlook, we plan to apply these ESE measurements to $\sqrt{s_{\rm NN}} = 200~{\rm GeV}$ Au+Au collision datasets to further investigate these trends.

\section*{Acknowledgments}

This material is based upon the work supported by JSPS KAKENHI Grant Number JP22K03648 and the QGP International Research Center, University of Tsukuba.

\bibliographystyle{elsarticle-num}
\bibliography{SQM2026_Kondo}

\end{document}